\documentclass[11pt,a4paper,onecolumn]{article}

\usepackage[comma,square,sort&compress,super]{natbib}

\usepackage{amsmath}
\usepackage{amssymb}
\usepackage{geometry}
\usepackage{graphicx}
\usepackage{float}
\usepackage[affil-it,auth-lg]{authblk}

\usepackage{color}
\usepackage[margin=1pt,font=small,labelfont=bf]{caption}

\title{\textbf{Optical positioning of single quantum dots in micropillar with $>$65\% extraction efficiency for on-demand quantum light sources}}
\author[1]{Shun-Fa Liu}
\author[1]{Yu-Ming Wei}
\author[1]{Rong-Ling Su}
\author[1]{Rong-Bin Su}
\author[2,3]{Ben Ma}
\author[2,3]{Ze-Sheng Chen}
\author[2,3]{\mbox{Hai-Qiao Ni}}
\author[2,3]{Zhi-Chuan Niu}
\author[1]{Ying Yu\thanks{yuying26@mail.sysu.edu.cn}}
\author[1]{Yu-Jia Wei\thanks{yujia.lisa.wei@gmail.com}}
\author[1]{Xue-Hua Wang}
\author[1,4]{Si-Yuan Yu}

\affil[1]{ State Key Laboratory of Optoelectronic Materials and Technologies, School of Electronics and Information Technology, School of Physics, Sun Yat-sen University, Guangzhou 510275, China}
\affil[2]{ State Key Laboratory of Superlattices and Microstructures, Institute of Semiconductors, Chinese Academy of Sciences, P.O. Box 912, Beijing 100083, China}
\affil[3]{ Synergetic Innovation Center of Quantum Information and Quantum Physics, University of Science and Technology of China, Hefei, Anhui 230026, China}
\affil[4]{Photonics Group, Merchant Venturers School of Engineering, University of Bristol, \mbox{Bristol BS8 1UB, UK}}

\begin{document}

\date{\today}

\maketitle

\begin{abstract}
  We report optical positioning single quantum dots (QDs) in planar cavity with an average position uncertainty $<$20 nm using an optimized two-color photoluminescence imaging technique. We create single-photon sources based on these QDs in determined micropillar cavities. The brightness of the QD fluorescence is greatly enhanced on resonance with the fundamental mode of the cavity, leading to an high extraction efficiency of 68\%$\pm$6\% into a lens with numerical aperture of 0.65, and simultaneously exhibiting low multi-photon probability ($g^{2}(0)$=0.144$\pm$0.012) at this collection efficiency.
\end{abstract}

\vspace{1cm}

Semiconductor quantum dots (QDs) are promising quantum light emitters of high quality single photons\cite{1lodahl2015interfacing,2aharonovich2016solid} and entangled photon pairs\cite{3gao2015coherent,4schwartz2016deterministic} which are important elements in scalable photonic quantum information processing\cite{5o2009photonic}. Resonant pulse excitation of a single quantum dot has generated pure, on-demanded and indistinguishable photons\cite{6he2013demand}. However, for QDs in bulk semiconductors, most emission photons from QDs will go back into the high refractive index material at air interface due to total reflection and cannot be extracted efficiently. By means of combining QDs with photonic microstructures\cite{7badolato2005deterministic,8moreau2001single,9maier2014bright,10pelton2002efficient,11claudon2010highly} that engineer the density of vacuum energies, spontaneous emission rate of QDs can be enhanced via the Purcell effect\cite{12purcell1946spontaneous} and extraction efficiencies can be increased remarkablely\cite{13ding2016demand,14unsleber2016highly,15somaschi2016near}. To improve the yield of useful devices, considerable efforts have been devoted to deterministically embed a single, pre-selected quantum emitter in various photonic structures to ensure that QDs match optical modes both in space and spectra\cite{7badolato2005deterministic,16gschrey2015highly,17dousse2008controlled,rev2Dousse2010,
18sapienza2015nanoscale}. Of these approaches, photoluminescence imaging at single-photon level is particularly attractive due to its positioning accuracy at a few tens of nanometers\cite{18sapienza2015nanoscale}, and its compatibility with high-resolution electron-beam lithography, which is typically used to pattern small features and can be process at room temperature.

Here we use the photoluminescence imaging technique developed in Ref.19 to determine the position of single QDs in planar distributed Bragg reflector (DBR) cavities with respect to fiducial alignment marks, with an average position uncertainty $<$20 nm. We also use this information to fabricate and demonstrate QD single-photon sources in micropillar cavities. Fine tuning of the QD line into the cavity resonance is obtained at temperatures ranging from 4 K to 40 K with a device yield of approximately 45\% in 47 devices. The device simultaneously exhibits high collection efficiency of 68\%$\pm$6\% into a lens with numerical aperture of 0.65, and low multiphoton probability ($g^{2}(0)$=0.144$\pm$0.012) at this collection efficiency.

\begin{figure}[htbp]
\centering
\includegraphics[width=\textwidth]{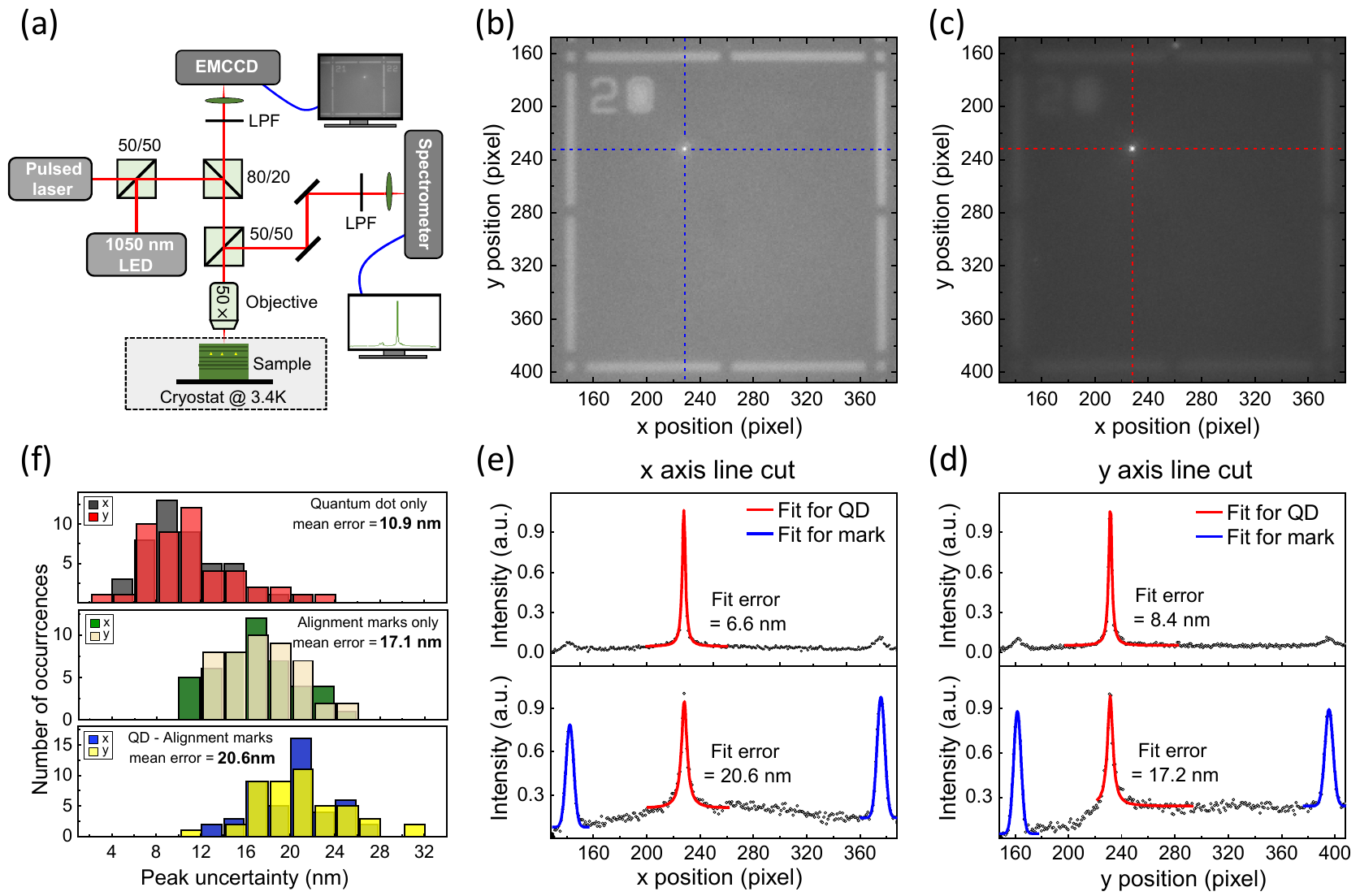}
\caption{(a) Schematic of the micro-photoluminescence and two-color photoluminescence imaging setup.
(b-e) Method to acquire the relative position of the QD:
(b) EMCCD image of the alignment marks when focusing on the surface.
(c) EMCCD image of the photoluminescence from a single QD when focusing on the QD layer that is at the center of $\lambda$-GaAs cavity ($\approx$1.85 $\mu m$ below the surface).
(d-e) \emph{x}(\emph{y}) axis line cut along the horizontal(vertical) dot line in (b) and (c), showing the QD emission, light intensity reflected by metallic marks. Herein, the Lorenz fit (red lines) and Gaussian fits (blue lines) are used to determine the location of the QD and the center position of alignment mark, respectively.
The positions are then translated from a pixel value on the images to a distance on the sample by counting the number of pixels between two nearby marks with known distance.
(f) Histograms of the uncertainties of the QD and alignment mark positions and QD-alignment mark separations (47 images). The uncertainties represent one standard deviation values determined by a nonlinear least squares fit of the data.}
\label{fig 1}
\end{figure}

The investigated sample consists a single layer of low density In(Ga)As QDs grown via molecular beam epitaxy and located at the center of a $\lambda$-thick GaAs cavity surrounded by two Al$_{0.9}$Ga$_{0.1}$As/GaAs Bragg mirrors with 12 (25) pairs. A silicon delta-doping was introduced 10 nm above the QD layer to stochastically charge the single QDs with an excess electron. The deterministically positioned QD-in-micropillar structures are processed by two-color photoluminescence (PL) imaging (Fig.\ref{fig 1}(a)) combined with standard electron-beam lithography. First, we select the QDs with their emission wavelengths near cavity mode by collecting their emissions with a microscope objective (NA= 0.65) into a grating spectrometer. Subsequently, spatial selection is achieved by imaging the QD positions with respect to alignment marks, which is incorporated into the same micro-PL set-up. This step ensures the positions of the QDs at the maximum of the pillar fundamental mode.

In detail, an array of Ti/Au metal alignment marks is fabricated on the surface of the DBR planar cavity structure through a standard lift-off process. The sample is then inserted in an optical microscopy cryostat (Montana, T=4 K-300 K) mounted on a motorized positioning system with piezo-electric actuators. The micro-PL and two-color PL imaging configurations are shown in Fig.\ref{fig 1}(a). A 800 fs pulsed laser with tunable wavelengths from 750 nm to 1040 nm and a 79.3 MHz repetition rate is used to give rise to a PL emission from the QDs, while a 1050 nm light emitting diode (LED) with a power of $\approx$2 mW is simultaneously used to illuminate the alignment marks, whose illumination wavelength is out of the stopband (870-980 nm) of our Bragg mirrors to regain contrast in the image. The laser beam was focused onto a selected QD with the laser spot of $\approx$1.5 $\mu$m. Reflected light and fluorescence from the sample go back through the 50/50 and 80/20 beam splitters and are imaged onto an Electron Multiplied Charged Couple Device (EMCCD). A 900 nm long-pass filter (LPF) is inserted in front of the EMCCD camera to remove reflected excitation light. The microscope objective is focused on the QD layer at the center of $\lambda$-GaAs cavity ($\approx$1.85 $\mu$m below the
surface) when imaging the fluorescence from the QDs, while imaging of the alignment marks is done by focusing on the planar surface of the structure to ensure its positioning accuracy.

Representative images of the alignment marks and QD photoluminescence at the different focal depths are shown in Fig.\ref{fig 1}(b) and \ref{fig 1}(c). When focused on the planar surface, as shown in Fig.\ref{fig 1}(b), a circular bright spot and related alignment marks are clearly visible, which represents the emission from one single QD within an $\approx$60 $\mu$m$\times$60 $\mu$m field of view. Orthogonal line cuts of the alignment marks are fitted with Gaussian functions using a nonlinear least squares approach, determining their centre positions with an typical uncertainty of $\approx$18.4 nm (Fig.\ref{fig 1}(d,e,f)). While focusing on the QD ($\approx$1.85 $\mu$m below the surface), as shown in Fig.\ref{fig 1}(c), the circular spot becomes optimally focused at the cost of fading the alignment marks, with the extracted peak of \emph{x}-positions with one standard deviation uncertainty as low as \mbox{6.6 nm}, much better than that of 20.6 nm when focused on the planar surface (Fig.\ref{fig 1}(e)). Here, the exposure time of EMCCD is set at 0.1 s to reduce sample drift during images acquisition. Furthermore, histograms of the measured values in Fig.\ref{fig 1}(f) show that the mean uncertainties in the quantum dot, alignment mark, and the QD-alignment mark separation are 10.9 nm, 17.1 nm, and 20.6 nm, respectively. Thus this two-color PL imaging technique allow us to determine the QD position by pointing the maximum of the QD emission according to the two-dimensional alignments marks.

\begin{figure}[hbtp]
\centering
\includegraphics[width=\textwidth]{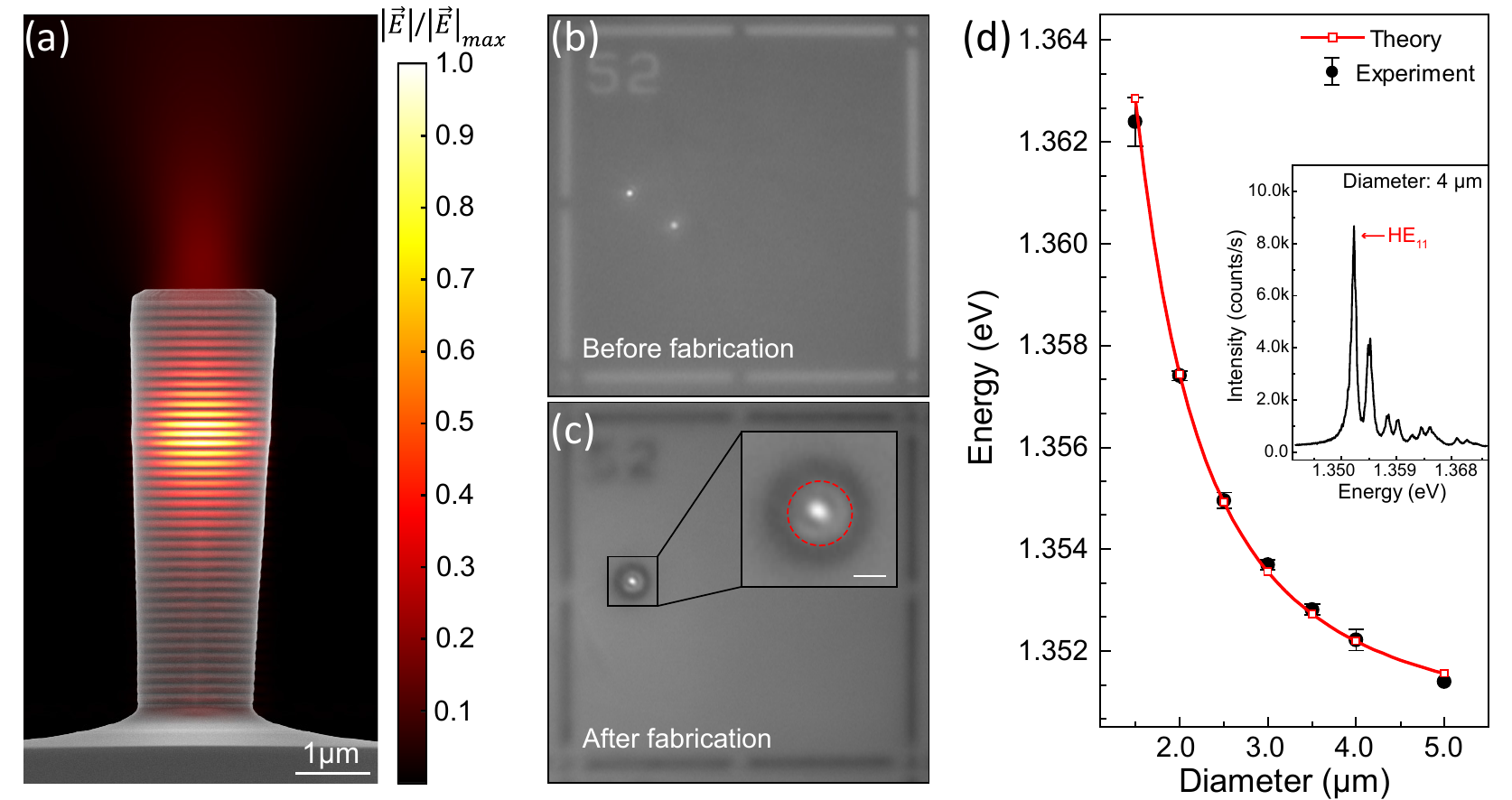}
\caption{(a) Scanning electron microscopy (SEM) image of a typical pillar with a diameter of 2  $\mu$m, along with the normalized electric field intensity distribution $|\vec{E}|$  calculated by 3D-FDTD method.
(b-c) Photoluminescence images of a 4 $\mu$m diameter micropillar with a single quantum dot in the center before (b) and after (c) fabrication. Scale bar represents 2 $\mu$m.
(d) The measured energy (black dots with error bars) of the fundamental mode (HE$_{11}$) for the pillar cavities as a function of the designed diameter, which are well described by theory according to \mbox{equation (\ref{eq1})} plotted in red line. Inset is a typical experiment cavity mode of a micropillar with a diameter of 4 $\mu$m acquired by raising the excitation power.}
\label{fig 2}
\end{figure}

After selecting the QD with desired photon energy (around cavity mode) and accurately determining its position, the pillar radius (R) is carefully designed according to the deviation of the emission frequency of the QD from the cavity mode to achieve spectral matching, as the energy of the pillar fundamental mode increases when the radius decreases \cite{19gerard1996quantum,20reitzenstein2010quantum}. And then, typical micropillar cavities are fabricated. The sample is first spin coated with a negative tone electron beam resist (HSQ fox15); The resist is exposed using a VISTEC EBPG5000 ES electron-beam lithography (EBL) system at 100 kV; Followed by the exposure and development process, the mask pattern of the pillar with a certain diameter is transferred into the sample via an inductively-coupled plasma reactive ion etching system (ICP-RIE, Oxford Instrument Plasmalab System 100 ICP180). A scanning electron microscopy (SEM) image of a typical pillar with a diameter of 2 $\mu$m is presented in Fig.\ref{fig 2}(a), which is superimposed with the normalized electric field intensity distribution ($|\vec{E}|$) calculated by 3D-FDTD method. Fig.\ref{fig 2}(b) and \ref{fig 2}(c) shows a representative photoluminescence image of the devices before and after fabrication, indicating a QD emission is just in the center of a micropillar structure. Fig.\ref{fig 2}(d) presents the measured and theoretical energy of the fundamental mode for the pillar cavities as a function of the designed diameter. The black circles represent the experiment cavity modes of different diameters acquired by raising the power of excitation laser, which is well matched to the theory result (red line) according to equation\cite{20reitzenstein2010quantum,21gutbrod1999angle}:
\begin{equation} \label{eq1}
E=\sqrt{E_{2D}^{2}+\frac{\hbar^2 c^2}{\epsilon}\frac{\chi_{n_{\varphi},n_{r}}^2}{R^2}}
\end{equation}
Where $E_{2D}$ is the resonance of the planar cavity, $\chi_{n_{\varphi},n_{r}}$
represents the $n_{r}^{th}$ zero of the Bessel function
$J_{n_{\varphi}(\frac{\chi_{n_{\varphi},n_{r}}}{R})}$, and R is the radius of
the pillar. For the fundamental HE$_{11}$ mode, the quantum numbers
($n_{\varphi}, n_{r}$, 0) is (1, 0, 0), and $\chi_{1,0}$ equals to 2.4048 here. By selecting appropriate pillar diameter for QD with different emission energy, we achieve a device yield of 45\% in 47 devices in matching the emission wavelength between QD and fundamental mode in the range of 4 K to 40 K. The deviation from an ideal fabrication process is mainly due to the large diameter interval of 0.5 $\mu$m, the slightly shifts of QD emissions during heating and cooling for several times (Fig.\ref{fig 4}d) or within the etching processes that change the strain environment of the QDs, which are also found in Ref.14.

\begin{figure}[phtb]
\centering
\includegraphics[width=8.5cm]{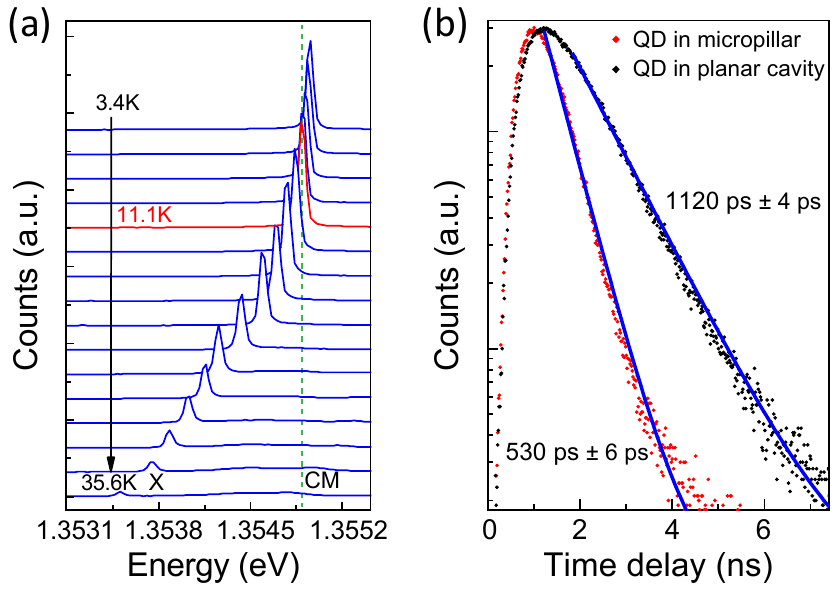}
\caption{(a) Temperature dependent spectra of a micropillar with a diameter of 2 $\mu$m, a strong enhancement on spectral resonance between fundamental mode (FM) and QD due to the Purcell effect is observed at T=11.1 K. (b) Time resolved measurements of the QD before fabrication (in planar structure) and in the micropillar cavity which reveal a Purcell factor of $F_{p}=2.1\pm0.3$.}
\label{fig 3}
\end{figure}

\begin{figure}[htbp]
\centering
\includegraphics[width=\textwidth]{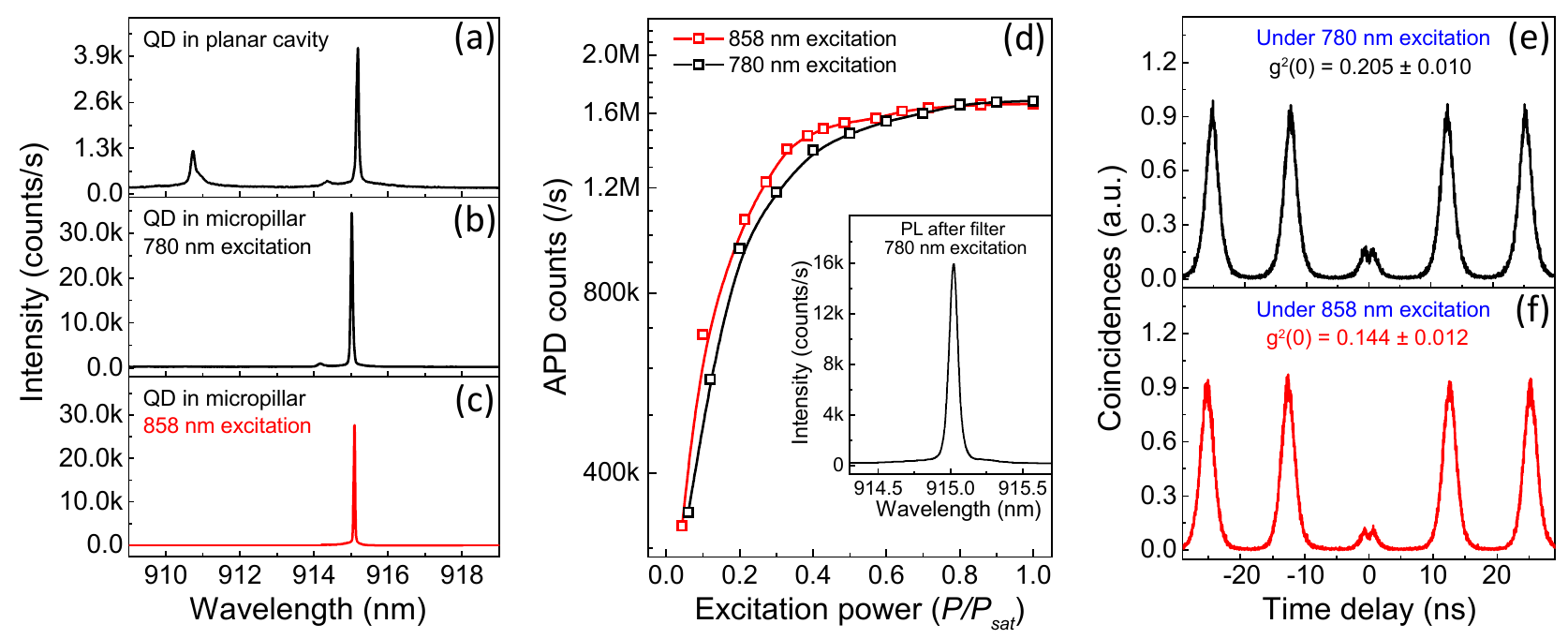}
\caption{(a-b) PL spectra of a single QD in a micropillar with a diameter of 2 $\mu$m before (a) and after (b) fabrication under non-resonant, 780 nm pulsed excitation. (c) PL spectrum of the QD in micropillar under 858 nm pulsed excitation. (d) Detected fluorescent counts of the same QD as a function of  the normalized pulse laser power $P/P_{sat}$ under 780 nm (black) and 858 nm (red) pulsed excitation, here $P$ and $P_{sat}$ represent to the excitation and saturation power. The inset shows a spectrum after a longpass filter and a narrow band filter with a bandwidth of 1 nm. (e-f) Intensity-correlation histogram obtained using a Hanbury Brown and Twiss-type set-up under 780 nm (e) and 858 nm (f) pulsed excitation.}
\label{fig 4}
\end{figure}

Now we turn to characterize the emission produced by the optically positioned QD within a micropillar with a diameter of 2 $\mu$m. The extracted Q factor of fundamental mode in our investigated device is 1438.47$\pm$1.51. A typical temperature dependent micro-PL is presented in Fig.\ref{fig 3}(a). A strong enhancement on spectral resonance between fundamental mode (FM) and QD due to the Purcell effect is observed at T=11.1 K. Time-photoluminescence measurements are carried out to determine the Purcell enhancement of the system. The spontaneous emission decay of the QD before fabrication (in planar structure) and in the micropillar cavity are shown in Fig.\ref{fig 3}(b). The single exponential fits of the decay curves indicate a lifetime of 530$\pm$6 ps for the QD in the micropillar cavity and a lifetime of 1120$\pm$4 ps for the QD in the planar structure, corresponding to a Purcell enhancement of the spontaneous emission rate by a factor of 2.1$\pm$0.3.

\begin{table}
\centering
\caption{\label{tab1}Experimental set-up calibration.}
\begin{tabular}{cccc}
 \hline
 \hline
 &Transmission&\text{Error bar}\\ \hline
Optical window&0.929&$\pm$3.0\%\\ \hline
50$\times$ microscope objective&0.787&$\pm$3.0\%\\ \hline
50/50 beam splitter&0.490&$\pm$3.0\%\\ \hline
50/50 beam splitter&0.490&$\pm$3.0\%\\ \hline
Silver mirror&0.956&$\pm$3.0\% \\ \hline
A 920 nm narrow band filter/&0.568&$\pm$2.0\%\\
a 900 nm long pass filter& &\\ \hline
A coupling lens&0.960&$\pm$3.0\%\\ \hline
Single-photon detector efficiency&0.300&$\pm$5.0\%\\ \hline
Overall detection efficiency&0.027&$\pm$9.1\%\\
 \hline
 \hline
\end{tabular}
\end{table}

To prove the brightness of this optical positioned QD in micropillar structure, we determine both the collection efficiency and the second order autocorrelation function at zero delay $g^2{(0)}$ when the QD emission is saturated. Fig.\ref{fig 4}(a-c) presents a PL spectrum of a single QD before and after fabrication under non-resonant (Fig.\ref{fig 4}(a)), 780 nm pulsed excitation (Fig.\ref{fig 4}(b)), and 858nm pulsed excitation (Fig.\ref{fig 4}(c)). Only the emission line which has a central wavelength (915.01 nm) within cavity mode appears with bright luminescence. In order to get pure QD fluorescence, a narrow band filter with a bandwidth of 1 nm is inserted into the collection arm of the confocal optical path. The inset in Fig.\ref{fig 4}(d) shows a spectrum after filtering in which only one peak remains. Fig.\ref{fig 4}(d) shows the detected fluorescent counts on a silicon single-photon detector as a function of normalized pulse laser power, achieving of 1,679,000 counts/s. To deduce the corresponding number of photons collected per excitation pulse in the first lens, we calibrate all the optical components of the detection path, as shown in Table \ref{tab1}. We estimate the total transmission rates of optical set-ups as $(2.7\pm$0.24)\%, where the uncertainty is based on the spread of transmission values measured for the optical components, and represents a one standard deviation value. To verify that these photons are true single photon, namely only one photon is generated when QD is driven by one laser pulse, we carried out an intensity-correlation measurement at saturated pump power density of 24 W/cm$^{2}$. The result is displayed in Fig.\ref{fig 4}(e). Although there is a dip at zero time delay which indicates only one photon generation at a time, two obvious small peaks around zero time delay lead to a $g^{2}(0)$ of 0.205$\pm$0.010. In these measurements, the proper mode-locking of the pulsed laser was carefully checked. These observations are not unique and occur in a similar way on a multitude of dots on this sample or other samples grown using the same MBE system \cite{ref1Yu2016}. We attribute these two peaks to a recapture process with assistance of trapped states in the QD sample\cite{ref2Dalgarno2008,ref3Aichele2004}. The carriers can be trapped in these states first for a certain time and after that there is a recapture process from trapped states into the QD following the initial recombination\cite{ref1Yu2016,ref2Dalgarno2008,ref3Aichele2004,ref4Nguyen2013}. To remove the effect of re-excitation, we multiply the total flux with $\frac{1}{1+g^2(0)}$ and get a pure single photon flux. Thus we estimate extraction efficiency that is the percentage of generated single photons collected into the first objective lens (NA=0.65) as 65\%$\pm$6\%. To obtain high pure combined with high brightness, we study the QD emission under pulsed 858 nm pulsed excitation (near wetting layer). There is only one peak left in the spectrum as shown in Fig.\ref{fig 4}(c). The power dependent fluorescent counts and the intensity autocorrelation measurement presented in Fig.\ref{fig 4}(d) and Fig.\ref{fig 4}(f) indicate a maximum of 1,657,000 counts/s with $g^{2}(0)$ of 0.144$\pm$0.012 at saturated pump power, revealing an extraction efficiency of 68\%$\pm$6\%.

In conclusion, we have realized positioning single QDs in planar cavity with respect to alignment marks with an average position uncertainty of $\approx$20 nm using an optimized two-color photoluminescence imaging technique. We have used this technique to create single-photon sources based on positioned QD in a micropillar cavity that simultaneously exhibit high brightness ($\eta$=68\%$\pm$6\%) and purity ($g^{2}(0)$=0.144$\pm$0.012). As a next step one could also implement a resonance fluorescence excitation to achieve highly indistinguishable on-demand photons. We believe this technique is an important step forward in the ability to create single QD micropillar devices due to its accurate positioning and effective mode coupling. The technique can be used in devices including strongly-coupled QD-microcavity systems\cite{24reinhard2012strongly,25hennessy2007quantum}, and orbital angular momentum modes (OAM) from quantum light sources\cite{26li2015orbital}, which is very encouraging for the implementation of integrated quantum dot based quantum circuits\cite{2aharonovich2016solid}.

\vspace{50pt}

\begin{Large}
\noindent
\textbf{Acknowledgments}
\end{Large}

\vspace{10pt}

The authors wish to thank Lin Liu, Li-Dan Zhou, Chun-chuan Yang, Zhi-Chao Nong for technical
assistance in microfabrication, as well as Xiong Wu, Ming-Bo He for valuable discussions.
This work is supported by the Natural Science Foundation of Guang-dong
Province (201676
12042030003), and the Specialized Research Fund for the Doctoral Program of
Higher Education of China (20167612031610002), National Key Basic Research Program of China
(2013CB933304), National Natural Science Foundation of China (91321313, 61274125).

\bibliographystyle{unsrt}

\end{document}